# Sub-nanometer-thick native $sp^2$ carbon on oxidized diamond surfaces

Ricardo Vidrio[1], Cesar Saucedo[2], Vincenzo Lordi[3], Shimon Kolkowitz[5], Keith G. Ray[3], Robert J. Hamers[2], Jennifer T. Choy[1,4]

[1]1500 Engineering Dr, Madison WI 53706, Department of Nuclear Engineering and Engineering Physics, University of Wisconsin-Madison

[2]1101 University Ave, Madison WI 53706, Department of Chemistry, University of Wisconsin-Madison

[3]7000 East Ave, Livermore CA 94550, Lawrence Livermore National Laboratory

[4]1415 Engineering Dr, Madison WI 53706, Department of Electrical and Computer Engineering, University of Wisconsin-Madison

[5]Physics South Hall, Berkeley, CA 94720-7300, Department of Physics, University of California, Berkeley

## Abstract

Oxygen-terminated diamond has a wide breadth of applications, which include stabilizing near-surface color centers, semiconductor devices, and biological sensors. Despite the vast literature on characterizing functionalization groups on diamond, the chemical composition on the shallowest portion of the surface (< 1 nm) is challenging to probe with conventional techniques like XPS and FTIR. In this work, we demonstrate the use of angle-resolved XPS to probe the first ten nanometers of (100) single-crystalline diamond, showing the changes of the oxygen functional groups and the allotropes of carbon with respect to depth. With the use of consistent peak-fitting methods, the peak identities and relative peak binding energies were identified for $sp^2$ carbon, ether, hydroxyl, carbonyl, and C-H groups. For the oxygen-terminated sample, we also quantified the thickness of the $sp^2$ carbon layer situated on top of the bulk $sp^3$ diamond bonded carbon to be 0.4±0.1 nm, based on the analysis of the Auger electron spectra and D-parameter calculations. These results indicate that the majority of the oxygen is bonded to the $sp^2$ carbon layer on the diamond, and not directly on the $sp^3$ diamond bonded carbon.

## Introduction

Depending on the type of surface termination employed, single-crystal diamonds (SCDs) can be tailored for a wide variety of applications ranging from DNA sensing [1], radiation detection [2], and stabilizing resonating nanostructures [3]. For example, oxygen (O)-terminated SCD has been shown to be an effective Schottky barrier diode using selective growing procedures and careful nanofabrication methods [4], while hydrogen (H) -terminated SCD has shown promise towards becoming viable for the detection of deep-ultraviolet light [5]. Quite recently, SCDs with either O or Nitrogen (N)-terminated diamond show promise towards enhancing the spin and optical properties of near surface color centers, thus enabling the use of SCD surfaces for the next-generation of quantum engineering applications [6] [7] [8]. Although there is a plethora of literature available on the study of functionalized SCD surfaces [9] [10] [11] [12], analysis of the shallowest portion of the surface proves difficult with common materials characterization techniques. While Fourier Transform Infrared (FTIR) spectroscopy is capable of discerning O functional groups on diamond materials [13] [14] [15] [16] [17], it unfortunately is not surface sensitive, as the penetration depth is on the order of microns. Furthermore, diamond is one of the hardest materials on Earth [18], making milling or machining for other methods, such as scanning tunnel microscopy (STM) or Atom Probe Tomography (APT) difficult, particularly so for SCD.

Although the prevailing model for the SCD surface has consisted of solely $sp^3$ carbon (C) bonds which are directly bonded to surface functionalizations [19] [20] [21] [22], recent experimental work, via angle-resolved X-ray photoelectron Spectroscopy (ARXPS) has provided evidence for the existence of a superficial layer of $sp^2$ C which rests on the bulk $sp^3$ diamond bonded C [23]. Given the recent growth of diamond research across multiple areas of science, an encompassing model of the surface of diamond will be useful for devising effective surface preparation techniques and accurate analysis of device performance. .

In this work, we performed ARXPS on O-terminated (100) diamond grown using Chemical Vapor Deposition (CVD), which enabled us to identify the multi-layered chemical structure in the first 10 nm below the diamond surface. We begin by interpreting the C1s spectra of two different surface treatments of (100) single-crystalline CVD diamond, a hydrogen (H) -terminated and O-terminated sample at multiple photoelectron emission angles. We show that the respective terminations on the diamond become less pronounced as the photoelectrons emanate from the deeper $sp^3$ C. Furthermore, we also assign regions on the spectra with their corresponding chemical identities, as well as report relative binding energy peak values for the singly bonded C-O groups, carbonyls, $sp^2$ C, and C-H peaks. Next, we analyze the C Auger electron lineshapes (CKLL) and calculate the D-parameter to determine the evolution from the superficial amorphous $sp^2$ C layer to the bulk $sp^3$ C diamond for the O terminated sample. Given the considerably shorter mean free path for auger electrons in diamond compared to core electrons, our technique probes a much shallower region of the diamond surface. This, combined with the stage tilting capabilities from ARXPS allows us to probe sub nanometer depth information from the SCD sample, giving us an unprecedented amount of information of the molecular nature of the C responsible for O-terminations. We conclude that the $sp^2$ C layer comprises only the first 0.4 nm of the surface. While prior studies [24] [25] [26] have reported on the presence of $sp^2$ C on O-terminated diamond surfaces, this finding elucidates the bonding characteristic of this native $sp^2$ C layer, namely that the O content is bonded to the $sp^2$ C and not directly on the bulk $sp^3$ C..

Experimental methods

All diamond samples used here are type IIa (100) SCD grown using CVD purchased from Element 6, with a surface roughness specified from the manufacturer polished to within < 30 nm. Two different surface treatments were studied, the O-termination and H-termination. The O-termination was accomplished by taking as-received diamond samples and performing a tri-acid clean bath, which consists of a 1:1:1 volumetric mixture of perchloric, sulfuric, and nitric acid at a constant temperature of 450°C, which results in a diamond surface of at least 4.84% O1s atomic percentage [26]. The H-terminated sample was first tri-acid cleaned and then put in a hydrogen plasma chamber to induce an effective H-termination in which no O1s or OKLL signal was observed from the XPS survey scan. All XPS peak fitting and atomic quantification was performed on CasaXPS software [27]. Analysis of the CKLL spectra was performed on python with the use of a modified form of the Whittaker filter, known as the Whittaker-Eilers smoothening method [28] [29].

To perform ARXPS a PHI versa probe 5000 was utilized and multiple scans were collected, including a survey scan, a narrow scan of the C1s spectra, a narrow scan of the O1s spectra, and a narrow scan of the CKLL. The flood gun was used at all times during the measurements to mitigate the effects of peak shifting. The PHI versa probe 5000 comes equipped with stage tilting capabilities which is capable of changing the angle at which the photoelectrons are emitted from the sample, which allows for the study of the photoelectron spectra at different depths. Shown below on Figure 1a is a schematic which illustrates the angle-tilting capabilities in the PHI versa probe 5000. The angle-tilting in this experiment progressed from a minimum angle of 10°, which represents the photoelectrons coming from the shallowest portions of the diamond, towards a maximum angle of 90°, in which the ejected photoelectrons come in from the bulk of the diamond. All angles come with an experimental uncertainty of 6˚, as the aperture was not included in the ARXPS setup due to the meager signal-to-noise (SNR) that was observed with the photoelectron spectra.

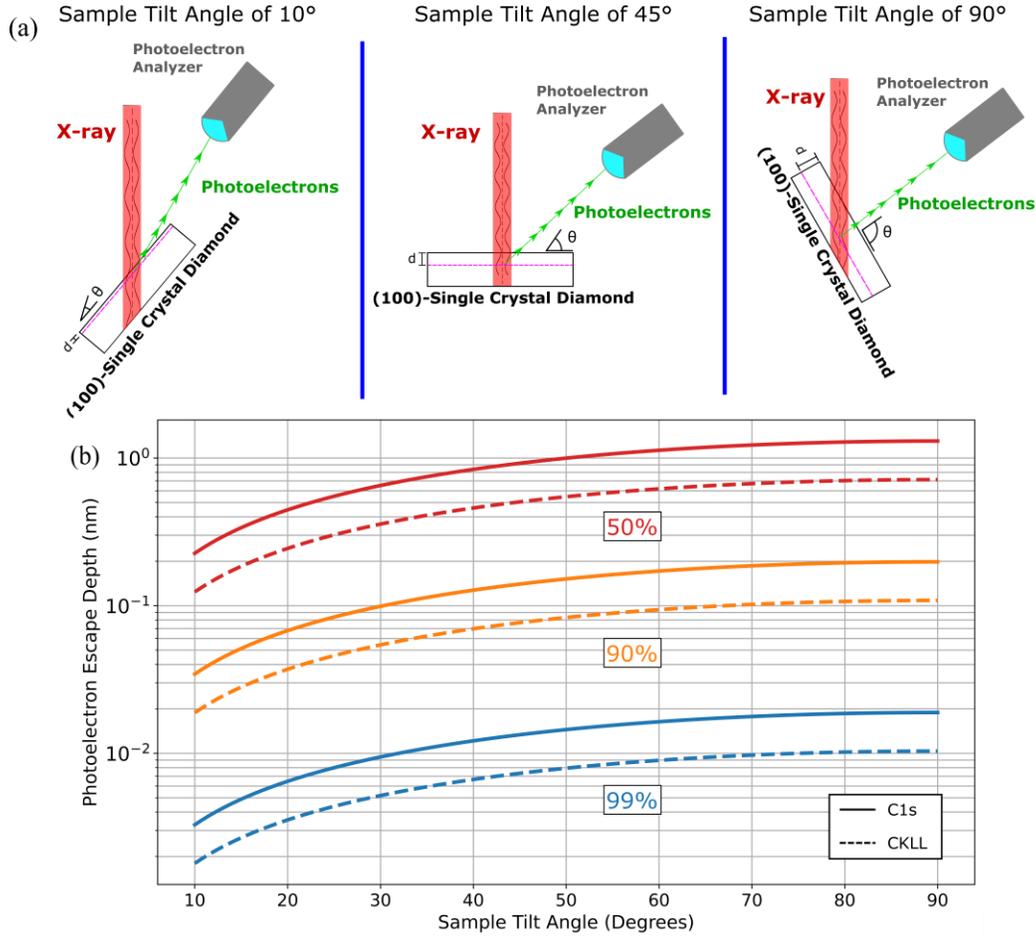

*Figure 1: (a) Experimental setup of the diamond sample within the PHI Versa Probe 5000 (b) Photoelectron escape depth as a function of sample tilt angle shown at three different escape probabilities: 50, 90, and 99%.*

To estimate the escape depths of the photoelectrons from the diamond sample, the photoelectron escape probability equation was used, which is shown on Equation 1.

*Equation 1*

$$P(z, \theta) = e^{\frac{-z}{\lambda \cdot \sin(\theta)}}$$

Here, $z$ represents the depth of the diamond in nm, $\lambda$ represents the inelastic mean free path of the photoelectron, and $\theta$ is the angle at which the stage is tilted. Alternatively, this equation can also be written in terms of the photoelectron escape depth at some constant $P(z)$ and plotted as a function of angle with respect to depth. Figure 1b represents the photoelectron escape depth as a function of sample tilt angles at three constant escape probabilities, at 50%, 90%, and 99%. Since we are analyzing two different allotropes of carbon within the same sample, two $\lambda$ values are plotted on Figure 1b, designated as the solid and dashed lines. Prior works in the surface analysis of SCD have shown evidence of a native $sp^2$ C layer that lies on the bulk $sp^3$ C diamond bonds [24] [23], but a detailed study regarding the specific allotropic information of this $sp^2$ C layer is lacking. It is most likely that the native $sp^2$ C layer is a form of amorphized C composed of mostly $sp^2$ C, as a past study on CVD grown diamond samples exposed to ns laser pulses and then subsequently treated with a $H_2SO_4$:$KNO_3$ acid mixture at 200°C for 30 minutes showed evidence of amorphization through analysis of Electron Energy Loss Spectroscopy (EELS) and Scanning Tunneling Electron Microscopy (STEM). The dotted line represents the $\lambda$ for the photoelectrons emanating from the amorphous $sp^2$ C layer at the relevant CKLL kinetic energies of 263 eV, which we take to be as 1.03 nm [30]. The solid line represents the $\lambda$ value for the photoelectrons for diamond, at a kinetic energy of 1202 eV, which corresponds to the peak location of the C1s spectra at a value of 1.88 nm [31].

Figure 1b highlights the surface sensitivity of the XPS method, as both the C1s and CKLL electrons are 50% likely to originate from roughly the first 1.3 nm of the surface throughout all angles surveyed during the experiment. Furthermore, throughout all escape probabilities we can observe that there is a depth mismatch between the depths that are sampled between the C1s and CKLL spectra, as the CKLL spectra is probing shallower depths when compared to the C1s spectra, due to the differing $\lambda$ values. This means that the CKLL electrons are roughly twice as surface sensitive as their C1s counterparts.

Results and Discussion

*Comparison of H-terminated and O-terminated surfaces*

Shown below on Figure 2 is a comparison of the C1s spectra for the H and O terminated diamond at different sample tilt angles. The arrows point to approximate peak regions corresponding to the unique chemical species which have been identified in prior works [7] [26] [32] [33] [34] and have been successfully deconvoluted using multiparameter peak-fitting methods described elsewhere [26] [35] [36] [37] [38] [39]. The only exception for the peak-fitting procedures taken from [26] is the constraint for the full width half maximum (FWHM) for the sp$^3$ C peak has been modified to within 0.6 to 0.86 eV, as the FWHM for the sp$^3$ C was found to have a noticeable variation due to the presence of chemical groups adjacent to the sp$^3$ C peak, which was more pronounced at the lower angles. FWHM values for all other chemical groups remained at 1.30 to 1.60 eV, as mentioned prior in [26].

For the purposes of a meaningful comparison for both diamonds across varying sample tilt angles, data are shown in terms of the relative binding energy (RBE), with all spectra shown relative to the bulk sp$^3$ C peak. A discussion regarding the origin of the peak shifting is reserved in the supplementary section of this work. In comparing both Figure 2a and Figure 2b we note the presence of the bulk sp$^3$ C peak which is always the most preeminent of all the spectral features across both samples for all measured angles. This in turn makes the sp$^3$ C peak useful as a reference across both data sets, and we thus define the sp$^3$ C peak at a RBE of 0 eV. O-terminated diamond (Figure 2b) displays more variations in surface chemistry when compared to the chemical homogeneity of the H-terminated diamond. Consistent with prior works on O-termination on SCD, we observe the presence of four chemical groups, sp$^2$ C, sp$^3$ C, and two oxygen functional groups denoted as C-O and C=O [32] [25] [7] [40] [26]. The singly bonded C-O peak comprises the ether and hydroxyl bonds on the diamond surface, while the C=O peak is indicative of the carbonyl bonding.

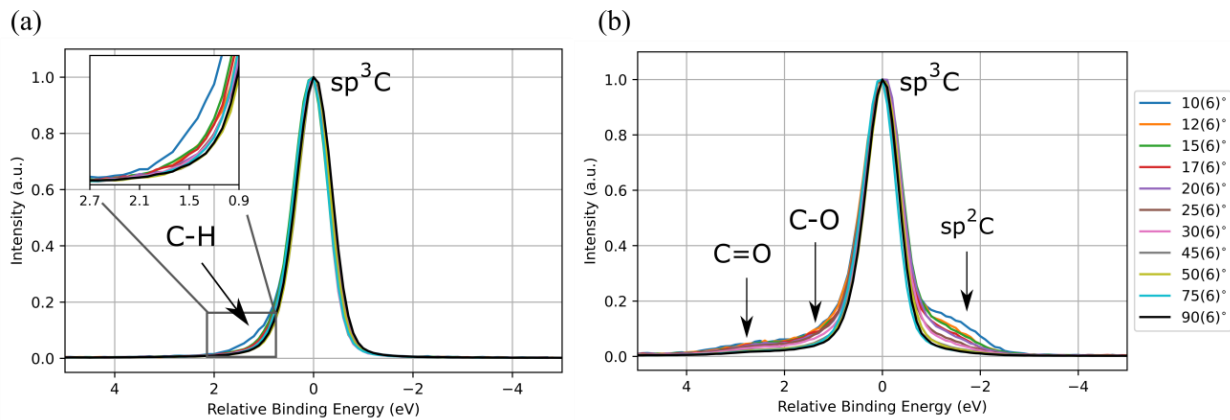

*Figure 2: (a) C(1s) spectra for H-terminated diamond at different sample tilt angles (b) C(1s) spectra for O-terminated diamond at different sample tilt angles.*

Both sets of spectra also change differently across varying sample tilt angles. The zoomed in plot of the H-terminated data in Figure 2a shows the sudden change in the C-H peak region as the sample tilt angle increases. The difference in the peak shape at the C-H region is readily apparent, as there is a wide gap present between the spectra taken at 10(6)˚ and the rest of the angles, with the data taken at 12(6)˚ and onwards showing a gradual decline in the C-H peak. By comparison, the O-terminated data shows a peak shape that exhibits more restrained changes as the spectra begins to penetrate deeper into the diamond. Based on these findings we confirm prior literature that has ascertained that an H-terminated diamond surface results in a chemically homogenized environment when compared to its O-terminated counterpart [41]. Furthermore, based on the rate at which the peak spectra changes with respect to angle, we ascertain that the H-terminated layer is lower than the superficial O and $sp^2$ C region.

Through established C1s peak fitting procedures, all chemical groups shown on Figure 2b were deconvoluted. However, peak-fitting the spectra from Figure 2a proved problematic as the location of the C-H and $sp^3$ C peak energies are too close in RBE value, which lends itself to ambiguity when attempting to perform a viable peak-fit. Therefore, only the H-terminated C1s spectra at 10(6)˚ was peak-fit, as the peak shape and position of the C-H region was more distinguishable when compared to the rest of the data. The values of the RBE for the two different surface treatments are tabulated below.

*Table 1: Relative binding energy values for all chemical species identified in both H and O terminated diamond samples.*

| H-terminated Diamond | |
|---|---|
| Chemical Species | Relative Binding Energy (eV) |
| $sp^3$ C | 0.00 |
| C-H | 0.46 |
| O-terminated Diamond | |
| Chemical Species | Relative Binding Energy (eV) |
| $sp^3$ C | 0.00 |
| $sp^2$ C | -1.22 ± 0.06 |
| C-O | 1.05 ± 0.09 |

| C=O | $2.63 \pm 0.07$ |
|---|---|

Angle- Dependent CKLL spectra and D-parameter analysis The acquisition and analysis of the CKLL spectra have been useful in the carbon community to determine the $sp^2$ C content of materials through the calculation of the D-parameter, a value which represents the amount of $\pi$ electrons inherent in $sp^2$ C [42], and manifests as the difference between the minimum and maximum point in the kinetic energies of the first derivative of the CKLL spectra. As Lascovich et. al, showed through comparison of graphite, diamond, and amorphous carbon samples, the D-parameter tends towards values of approximately 22.5 eV for graphite, while for diamond this will decrease to roughly 14.0 eV [42] [43] [44]. In essence, $sp^2$ C rich materials will tend towards higher D-parameter values, while materials heavy in $sp^3$ C will exhibit a lower D-parameter. Using this, it is possible to calculate the $sp^2/sp^3$ C fraction in a material via linear interpolation of the D-parameter from a range of about 14.0 to 22.5 eV [43].

Although CKLL analysis have been performed on diamond-like materials, interpreting the $sp^2$ C content on bulk diamond samples with the D-parameter comes with certain challenges. As evidenced on Figure 1, the IMFP for energies relevant to the C1s and CKLL spectra are remarkably different, being 1.88 nm and 1.03 nm respectively. This corresponds to the CKLL photoelectrons being roughly twice as surface sensitive as the C1s electrons. This means for the same measurement, a direct comparison between the $sp^2/sp^3$ C fraction that was derived from the C1s peak-fitting with that of the $sp^2/sp^3$ C value that was calculated from the D-parameter would not be valid, as there is a mismatch in the depth information between both spectra. In fact, this is a limitation of the CKLL analysis technique that has been observed prior in microcrystalline diamond films [45].

However, given the extreme surface sensitivity of the CKLL photoelectrons, one can probe the shallowest of depths in SCD in a way that is not possible with analysis of the C1s spectra. One of the most relevant pieces of information that can be gleaned from this is the presence of a native $sp^2$ C layer that seems to be inherent on the surface of the SCD [24] [25]. The data on an estimate of this layer is sparse, but an approximate value based on experimental ARXPS data was calculated by assuming a three-layered model of the SCD surface consisting of a first layer comprised of C bonded to O in a 1:1 ratio, a second layer formed of solely $sp^2$ C, and finally the bulk $sp^3$ C diamond [23]. The non-diamond region of the SCD surface, the first and second layer, were calculated to be 0.089 nm and 0.27 nm respectively, making the cumulative amount of the non sp$^3$ C region as 0.33 nm.

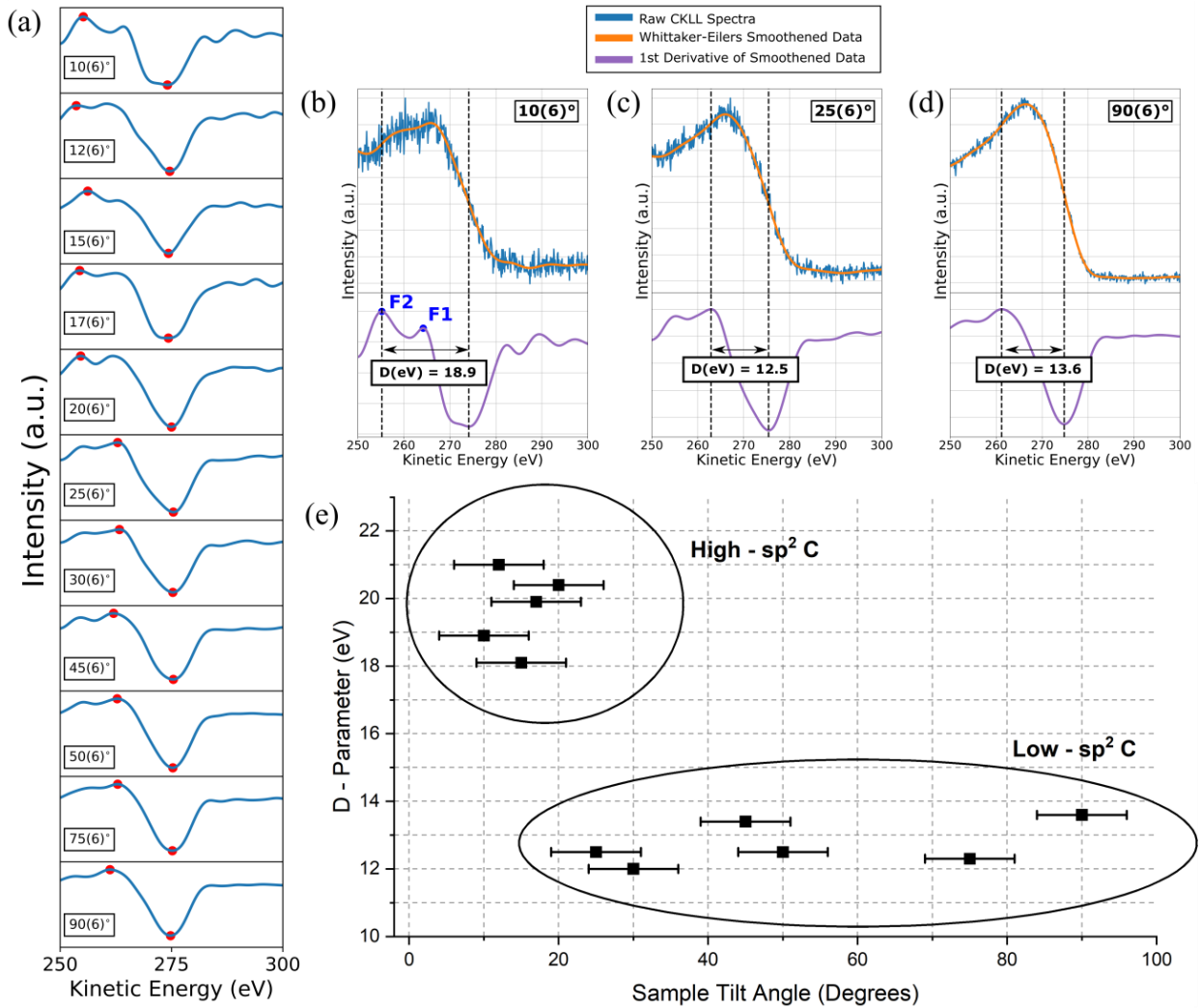

*Figure 3: (a) 1$^{st}$ derivative of smoothened CKLL spectra for all sample-title angles. (b) Raw CKLL spectra and 1$^{st}$ derivative of smoothened data shown with the calculated D parameter at 10(6)˚ (c) Raw CKLL spectra and 1$^{st}$ derivative of smoothened data shown with the calculated D parameter at 30(6)˚ (d) Raw CKLL spectra and 1$^{st}$ derivative of smoothened data shown with the calculated D-parameter at 90(6)˚ (e) Calculated D-parameter values plotted as a function of sample tilt angles. Also shown are the regimes of high sp$^2$ C and low sp$^2$ C which happens abruptly at 30(6)˚*

Here, we arrive at a value of the thickness of the non $sp^3$ C region without any assumptions of the carbon allotrope layering or the C:O ratio on the surface. Shown on Figure 3a are the plots for the first derivative of the smoothened CKLL raw spectra taken for the O-terminated SCD sample at all sample tilt angles. All raw spectra were first smoothened with a modified form of the Whittaker smoothening method, known as the Whittaker-Eilers smoother [28] [29] and then differentiated. The red points signify the highest and lowest points in the differentiated spectra, whose difference represents the D-parameter. These D-parameter values for all angles are then plotted on Figure 3e, from which we can see that there are both a high and low region of D values. The higher D value region represents the depth in the SCD layer with a high $sp^2$ C content which is predominant at $10(6)^{\circ}$ to $20(6)^{\circ}$. At $25(6)^{\circ}$ this value drops immediately to 12.5 eV and then is roughly constant at 13.0 eV for all subsequent angles.

Figure 3a illustrates the first derivative of the raw CKLL spectra that was smoothened by the Whittaker-Eilers smoothening method. The red points indicate the minimum and maximum points in the spectra, whose difference is in the D-parameter. A limitation in this data processing method is that the location in the maximum point can become hard to distinguish due to the smoothening algorithm. This discussion is reserved for the supplementary section of this work and is shown explicitly with data at an angle of $12(6)^{\circ}$.

Figure 3b to 3d show both the raw CKLL spectra and the smoothened first derivative data. Labeled on the subfigures are the maximum and minimum values in the differentiated spectra, shown as vertical dotted lines, and the D-parameter values, shown as the horizontal arrows between the dotted lines and labeled accordingly underneath each of the arrows. Three angles are shown in the subfigures, being $10(6)^{\circ}$, $25(6)^{\circ}$, and $90(6)^{\circ}$ respectively. Each of these angles represents different regions of carbon in the SCD layering. For example, at $10(6)^{\circ}$, we see two peaks labeled as F2 and F1 at 255.2 and 263.5 eV respectively. The F1 and F2 peaks have been observed in prior works with high $sp^2$ C materials, as both peaks were present in graphite samples [46] [45], as well as in amorphous C [30]. These two peaks are present within roughly 247 and 267 eV of the CKLL spectra. It is then the energy difference between the F2 peak and the minimum in the spectra that results in a D value indicative of a C sample with a content of at least 50% $sp^2$ C. Prior work has attributed the F2 and F1 peaks as representing the σ-σ and σ-π partial local electron densities respectively [47]. The presence of the F1 and F2 peaks at $10(6)^{\circ}$ indicate that the shallowest portion of the SCD is comprised of a high amount of $\pi$ bonds that are common in $sp^2$ C. Following $10(6)°$, we then notice that the location around the F2 peak starts to wane, and what used to be the F1 peak starts to shift slightly with respect to kinetic energy and increases in height. Note however, that this is a gradual process visible only throughout $10(6)°$ to $20(6)°$.

As shown on Figure 3c, by the time the sample tilt angle increases to $25(6)°$, the highest peak position has now shifted to 262.9 eV. This in turn yields a D-parameter value of 12.5 eV. All subsequent angles show similar behavior, as evidenced by Figure 3e. Shown on Figure 3d is the raw and smoothened first derivative of the CKLL spectra at a sample tilt angle of $90(6)°$ which represents the bulk diamond layer in the SCD. We can observe that the spectra has changed little since $25(6)°$, with a D-parameter of 13.6 eV. Therefore, the angles at $10(6)°$, $25(6)°$, and $90(6)°$ are representative of three distinct regions in the SCD layering. First an amorphous C region rich in $sp^2$ C bonds, second, the point at which the materials starts to shift from being composed of mostly $sp^2$ C to $sp^3$ C, and finally the bulk diamond, predominantly composed of $sp^3$ C bonds. From this we can discern that the amorphous $sp^2$ C layer extends within roughly $10(6)°$ to $20(6)°$, otherwise labeled as the *High-$sp^2$ C* region in Figure 3e. Past $20(6)°$ our D values start to drop and are centered around 13.0 eV. This region is labeled as the *Low-$sp^2$ C region*, comprised of angles between $25(6)°$ to $90(6)°$.

## *Calculation of the $sp^2$ C Layer*

With this information we can now estimate an upper bound of the depth of the amorphous carbon $sp^2$ C region on the SCD sample. We assume here that the D-parameter that is measured from our sample follows the same linear relation between D values and % $sp^2$ C content for other high D-parameter materials. Using this relation, we can relate the amount of $sp^2$ C to the photon electron escape probability using Equation 2.

*Equation 2*

$$\%sp^2C = \frac{\int_0^T e^{\frac{-z}{\lambda_{a-C}\cdot\sin(\theta)}}dz}{\int_0^T e^{\frac{-z}{\lambda_{a-C}\cdot\sin(\theta)}}dz + \int_0^\infty e^{\frac{-T}{\lambda_{a-C}\cdot\sin(\theta)}}\cdot e^{\frac{-z}{\lambda_{sp^3C}\cdot\sin(\theta)}}dz}$$

The left hand of the equation, % $sp^2$ C is the percentage of the $sp^2$ C content, obtained from the D-parameter [46]. On the right-hand side of the equation, we have a ratio of the integrated photoelectron escape probabilities. The numerator relates only to the integrated photoelectron escape probability from the sp$^2$ C layer, while the denominator is the integrated photoelectron escape probability from both the bulk diamond and sp$^2$ C layer.

The T value in the integration bounds represents the thickness of the amorphous sp$^2$ C layer, the $\lambda_{a\text{-}C}$ value is the IMFP for amorphous carbon, and $\lambda_{sp^3C}$ is the IMFP for diamond. Note the denominator is the total sum of the integrated photoelectron escape probability from the sp$^2$ C layer and the bulk diamond, with the integral from zero to infinity representing the total probability of a photoelectron exiting both the bulk diamond and the sp$^2$ C layer. The $e^{\frac{-T}{\lambda_{a\text{-}C}\cdot\sin(\theta)}}$ expression is equal to the photoelectron escape probability right at the sp$^2$ C layer, which is then multiplied by $e^{\frac{-z}{\lambda_{sp^3C}\cdot\sin(\theta)}}$ to yield the probability of a photoelectron exiting the diamond and amorphous C layer. Solving this equation for T then results in the following.

*Equation 3*
$$T(\theta, \%sp^2C) = -\lambda_{a-C}\cdot\sin(\theta)\ln\left(\frac{\lambda_{a-C}(\%sp^2C - 1)}{\lambda_{a-C}(\%sp^2C - 1) - \%sp^2C\cdot\lambda_{sp^3C}}\right)$$

The amorphous sp$^2$ C layer is then a function of the percentage of the sp$^2$ C content and the sample tilt angle. Here we assume that the values for both $\lambda_{a\text{-}C}$ and $\lambda_{sp^3C}$ are at a constant kinetic energy of 263.0 eV. Given a value of 20° (and an angle uncertainty of 6°), a D-parameter of 20.4 eV, corresponding to 74% of sp$^2$ C, we attain a sp$^2$ C layer depth of 0.4±0.1 nm.

The implications of a model of the single crystal diamond surface that consists firstly of an amorphized C layer means that the O functional groups are bonded to the sp$^2$ C and not directly to the sp$^3$ C. This is also evidenced by analyzing the relationships between the sp$^2$/sp$^3$ C fractions, the O1s atomic percentage, and the total O-functionalization area from the survey and C1s narrow scan spectra. Figure 4a illustrates the O1s atomic percentage plotted against the fractional amounts of sp$^2$/sp$^3$ C, where each data point has been color coded with respect to the color map shown on the right-hand side of the subplot to signify the change in sample tilt angle that each point corresponds to. The data show that at the shallower angles the O1s atomic percentage plateaus with respect to the amount of sp$^2$ C in the sample. As the XPS probes deeper into the sample, the O1s atomic percentage diminishes, as the diamond becomes more sp$^3$ C rich. These results are consistent with the total O content in the diamond being constrained mostly to the sp$^2$ C layer, and not necessarily the sp$^3$ C portion.

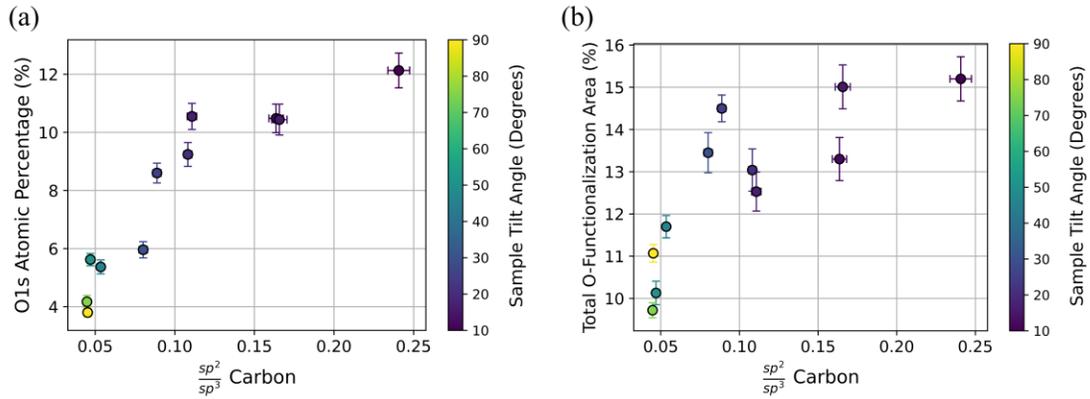

*Figure 4: (a) sp2/sp3 C fraction shown alongside O1s Atomic Percentage. (b) Total O-Functionalization area shown as a function of fractional amounts of sp2/sp3 C. The color bar on the righthand side of the plot represents the change in sample tilt angle for all data points on the figure.*

Figure 4b shows the fractional amount of $sp^2/sp^3$ C against the total O-functionalization area percentage, with each data point color coded to represent the corresponding sample tilt angle. At the shallower sample tilt angles, we see a sharp increase in the amount of total O-functionalization area percentage, that corresponds to increasing amounts of $sp^2$ C. Similar to how the O1s atomic percentage drops off in Figure 4a, we see a sharp decrease in O-functionalization as the sample transitions towards being $sp^3$ C rich. This means that the vast majority of the O-functional groups are present on the $sp^2$ C portions of the diamond. Taken together, Figure 4a and 4b illustrate that the O rich portions of the diamond tend to be constrained on the $sp^2$ C rich regions. The reader will note that values of O1s atomic percentage or total O-functionalization area percentage never fall down exactly to zero as the $sp^2$ C decreases. This is due to the fact that despite probing deeper into the diamond, some nonzero amount of photoelectrons still emanate from the superficial amorphous C layer, although most of them will correspond to the $sp^3$ C rich regions.

Conclusions

To summarize, this work represents a thorough study of the evolution of both the C1s and CKLL spectra with ARXPS at varying sample tilt angles at 10(6)° to 90(6)°. Analysis of the C1s spectra for an O-terminated sample reveals higher amounts of $sp^2$ C and functional groups present at the shallower angles that gradually wane as the photoelectrons begin exiting from the bulk diamond. The H-terminated sample reveals only the presence of $sp^3$ C and C-H bonds with no $sp^2$

C or functional groups. Through the use of peak-fitting techniques shown in prior works, we assign peak positions in terms of RBE values for all chemical species identified.

And finally, our analysis of the CKLL spectra reveals layering of different carbon allotropes, amorphous $sp^2$ C and $sp^3$ diamond C, through the calculation of the D-parameter at all sample tilt angles. By using the linear relation between the D-parameter and the percentage of the $sp^2$ C content, and the integrated photoelectron escape probabilities through both the $sp^2$ C and bulk diamond region, we calculate an upper estimate of the amorphous $sp^2$ C depth as 0.4±0.1 nm. This likely means that for O-terminated SCD, the O functional groups are bonded not directly to the $sp^3$ C diamond, but rather, they are bonded to the amorphous $sp^2$ C layer. By utilizing the results from the survey scan and the peak fitting from the C1s spectra, we observe that the total O content and O-functionalization area percentage is constrained to the $sp^2$ C rich portions of the sample. Together, these results signify that the O is bonded directly to the $sp^2$ C layer, and not directly on the $sp^3$ C region.

Future work should compare $sp^2/sp^3$ C values with the C1s and CKLL spectra by finding optimal angles in the ARXPS where the ratio of the photoelectron escape probability for the C1s spectra $P_{C1s}$ and the photoelectron escape probability for the CKLL spectra $P_{CKLL}$ are roughly identical, or $\frac{P_{C1s}}{P_{CKLL}} \approx 1$. Another issue that remains pertains to whether the O is bonded on top of the amorphous $sp^2$ C layer or encompasses the same region as the $sp^2$ C region. Subsequent studies should place emphasis on a comparison of the O1s and OKLL spectra alongside the C1s and CKLL spectra, and analyzing how it evolves with changing sample tilt angle. However, one will also have to keep in mind the relevant escape depths for the photoelectrons when analyzing O1s, C1s, OKLL, and CKLL data.

Acknowledgements

This work is supported by the U.S. Department of Energy, Office of Science, Basic Energy Sciences under Award #DE-SC0020313.

## Sub-nanometer thick native sp$^2$ carbon on oxidized diamond surfaces

Ricardo Vidrio[1], Cesar Saucedo[2], Vincenzo Lordi, Shimon Kolkowitz, Keith G. Ray[3], Robert J. Hamers[2], Jennifer T. Choy[1,4]

[1]1500 Engineering Dr, Madison WI 53706, Department of Nuclear Engineering and Engineering Physics, University of Wisconsin-Madison

[2]1101 University Ave, Madison WI 53706, Department of Chemistry, University of Wisconsin-Madison

[3]7000 East Ave, Livermore CA 94550, Lawrence Livermore National Laboratory

[4]1415 Engineering Dr, Madison WI 53706, Department of Electrical and Computer Engineering, University of Wisconsin-Madison

## Peak shifting effects

Peak-shifting effects were visible in the carbon photoelectron spectra for both H and O terminated diamond samples despite the flood gun being utilized. The peak-shifting has been seen in prior work for H-terminated diamond, as the C-H bonds contribute to upward band bending, which in turn reduces the barrier needed for electron emission from the sp$^3$ C peak position , but becomes particularly more noticeable here due to how the band structure changes as the photoelectrons start out emanating from the rich C-H layer at the lower angles and then transition towards the bulk sp$^3$ C at higher angles [44]. We record the maximum peak shifted across the H-terminated diamond sample measurements as 0.40 eV.

The origin of the peak shifts in the oxygen terminated sample most likely results from the effects of vertical differential charging (VDC) which stems from the layered chemical structure of the diamond sample, which features a prominent presence of electrically conductive sp$^2$ C in the shallower regions at lower angles, followed by the insulating sp$^3$ C in the bulk at higher angles. VDC effects have been observed in heterogenous materials that feature an insulating thin overlayer followed by a bulk conducting substrate, where in those works, peak shifting was noticeable even with the implementation of charge neutralization methods [45] [46] [47]. Here we see the same shifting behavior, albeit conversely with a thin electrically conductive layer followed by a bulk insulator. At the lower angles the material behaves like an electrical conductor, and therefore the charge neutralization effects of the flood gun would overcompensate for any surface charging effects, resulting in lower BE values for the bulk sp$^3$ C peak. Contrarily, as the x-rays penetrate deeper into the diamond, and the photoelectrons originate from the bulk sp$^3$ C region, the kinetic energies of the exiting photoelectrons will be slightly hindered as they depart from an insulating material, resulting in higher BE values. The maximum value in the peak shifts for the O-terminated sample was recorded as 0.30 eV.

## Assining maximum and minimum values in CKLL spectra

The reader will note that in the figure below at 12(6)˚ there are both a green and red point present at lower kinetic energies. The green point represents the highest value that was chosen for the calculation of the D-parameter, as it was found that the maximum (the red point to the left of the green) yielded a D value of 14.2 eV, which in turn was inconsistent with the rest of the data between 10(6) to 20(6)˚. This represents some of the inherent limitations of the D-parameter analysis technique, as it is imperative that one smoothens their CKLL raw spectra before differentiating, as the auger spectra is normally too noisy to just differentiate outright. Improperly applying a smoothening method to the CKLL spectra can lead to erroneous D-parameter values, via either under or over smoothening. In this case, it was found that at 12(6)˚ the height difference between the green and red point was only about 0.16 a.u. apart. Seeing as how the D-parameter at 12(6)˚ was the only value between 10(6) to 20(6)˚ to not be higher than 18 eV, it was determined that the green point was the true maximum value, with the other maximum value being present there only as an artifact of the smoothening method. After factoring in this change, the revised D-parameter was calculated to be 21.0 eV.

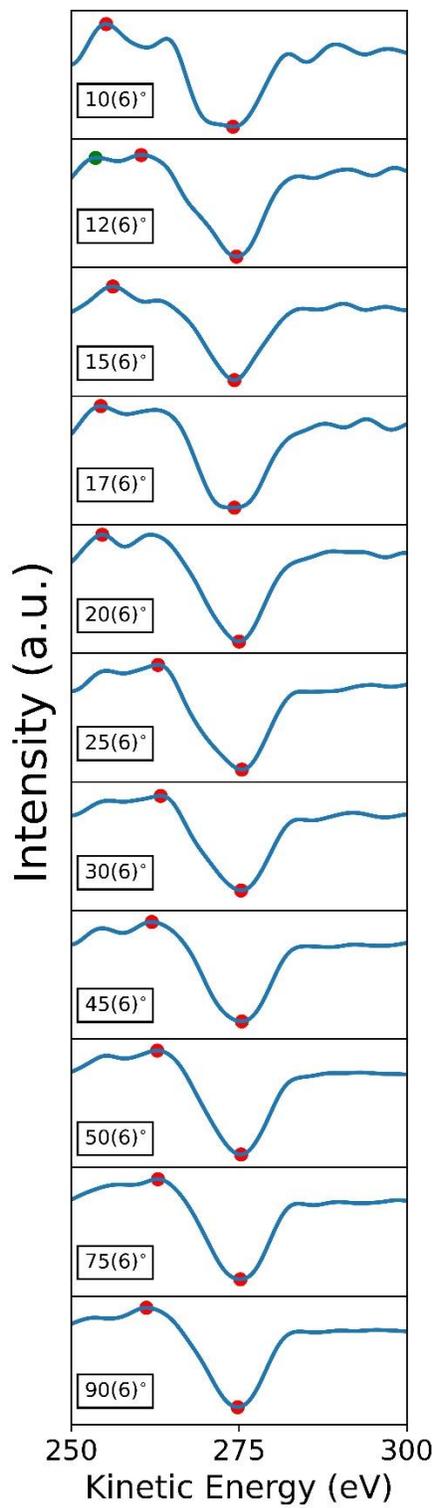